\documentstyle[aps,epsfig,twocolumn,floatfig]{revtex}

\newcommand{\up}[0]{\vert\uparrow\rangle}
\newcommand{\down}[0]{\vert\downarrow\rangle}
\newcommand{\uu}[0]{\vert\uparrow\uparrow\rangle}
\newcommand{\dd}[0]{\vert\downarrow\downarrow\rangle}
\newcommand{\ud}[0]{\vert\uparrow\downarrow\rangle}
\newcommand{\du}[0]{\vert\downarrow\uparrow\rangle}
\newcommand{\ket}[1]{\vert{#1}\rangle}
\newcommand{\ns}[2]{\vert\langle{#1}\vert{#2}\rangle\vert^2}

\begin{document}

\title{Deterministic entanglement of two trapped ions\cite{gov}}
\author{
Q.\ A.\ Turchette,\cite{email} C.\ S.\ Wood, B.\ E.\ King,
C.\ J.\ Myatt, D.\ Leibfried,\cite{didi}
W.\ M.\ Itano, C.\ Monroe, and D.\ J.\ Wineland
}
\address{
Time and Frequency Division, National Institute of Standards and Technology,
Boulder, CO 80303
}
\date{\today}
\maketitle

\begin{abstract}
We have prepared the internal states of two trapped 
ions in both the Bell-like singlet
and triplet entangled states. 
In contrast to all other experiments with entangled states of
either massive particles or photons, 
we do this in a deterministic
fashion, producing entangled states {\it on demand} without 
selection. 
The deterministic production of entangled states is a crucial
prerequisite for large-scale quantum computation.
\end{abstract}


Since the seminal discussions of Einstein, Podolsky, and Rosen,
two-particle quantum entanglement has
been used to magnify and confirm the peculiarities of quantum 
mechanics \cite{EPRBohmBell87}.  More
recently, quantum entanglement has been shown to be not purely of 
pedagogical interest, but also relevant to computation
\cite{Shor97Grover97a}, information transfer \cite{Barenco95},
cryptography \cite{Ekert91} and spectroscopy \cite{Bollinger96,Wineland98}.
Quantum computation (QC) 
exploits the inherent parallelism of quantum superposition and entanglement to 
perform certain tasks more efficiently than
can be achieved classically \cite{Ekert96bGrover98}.

Relatively few physical systems are able to approach the severe requirements
of QC: controllable coherent interaction between the quantum
information carriers (quantum bits or qubits), isolation 
from the environment, and high-efficiency interrogation of individual qubits. 
Cirac and Zoller have proposed
a {\it scalable} scheme utilizing trapped ions for  QC \cite{Cirac95}.
In it, the qubits are two internal states of an ion;
entanglement and computation are achieved by quantum logic operations
on pairs of ions involving shared quantized motion.  
Previously, quantum logic operations were demonstrated between a
single ion's motion and its spin \cite{Monroe95}; the requirements of QC
have been explored experimentally in related
cavity QED systems \cite{Turchette95Maitre97}. In this Letter, 
we use conditional quantum logic transformations to entangle
and manipulate the qubits of two trapped ions.

Previous experiments have studied entangled states of photons
\cite{Freedman72Fry76Aspect82Aspect82b,Shih88Ou88Ou92Kwiat95Tittel97Bouwmeester97} 
and of massive particles\cite{Lamehi76,Hagley97,Laflamme97}.  
These experiments rely in some way on {\it random processes}, 
either in creation of the entanglement,
as in photon cascades \cite{Freedman72Fry76Aspect82Aspect82b},
photon down-conversion \cite{Shih88Ou88Ou92Kwiat95Tittel97Bouwmeester97}
and proton scattering \cite{Lamehi76}, or in the random arrival times of
atoms in a cavity \cite{Hagley97}.  
Recent results in NMR of bulk samples have shown entanglement
of particle spins
\cite{Laflamme97,Chuang98Chuang98aCorey97Corey98}
but because pseudo-pure states are selected through averaging over 
a thermal distribution, the signal 
is exponentially degraded as the number of qubits is increased.  
All the above processes are {\it selectable} but are not
{\it deterministic} generators of entanglement. 
By deterministic, we mean that a known and controllable 
quantum state of (all of)
a given set of particles is generated
{\it at a specified time} \cite{Law97}.
Deterministic entanglement coupled with the ability to store entangled
states for future use is crucial for the realization of 
large-scale quantum computation.   
Ion-trap QC has no fundamental scaling limits; moreover, even the simple
two-ion manipulations described here can, in principle, be
incorporated into large-scale computing, either by coupling two-ion
subsystems via cavities \cite{Cirac97}, or by using accumulators
\cite{Wineland98}.

In this Letter, we describe the deterministic generation of a state
which under ideal conditions is given by
\begin{equation}
\vert \psi_e(\phi) \rangle  = \left[ \frac{3}{5} \du 
- e^{i\phi} \frac{4}{5} \ud  \right] 
\label{eq:estate} 
\end{equation}
where $\down$ and $\up$ refer to internal electronic states of each ion
(in the usual spin-$1/2$ analogy) and
$\phi$ is a controllable phase-factor.
 For $\phi = 0$ or $\pi$, $\ket{\psi_e(\phi)}$ is a good approximation to
the usual Bell singlet ($-$) or triplet ($+$) state 
$\vert \psi_B^\mp \rangle = [\du \mp \ud]/\sqrt{2}$ since
$\ns{\psi_B^-}{\psi_e(0)} = \ns{\psi_B^+}{\psi_e(\pi)} = 0.98$ 
\cite{Hill97}.  The fidelity of our experimentally generated state
described by density matrix $\rho^\pm$ is 
$\langle\psi_e(\pi,0)\vert\rho^\pm\vert\psi_e(\pi,0)\rangle\approx
\langle\psi_B^\pm\vert\rho^\pm\vert\psi_B^\pm\rangle
\approx 0.70$, so that for all practical purposes, we can 
consider $\rho^\pm$ to be an approximation to the Bell states.
We describe a novel means of differentially addressing  each ion 
to generate the entanglement and a state-sensitive
detection process to characterize it.

The  apparatus is described 
in Ref. \cite{King98}.  We confine $^9$Be$^+$ ions in an elliptical
rf Paul trap (major axis $\approx 525 \mu$m, aspect ratio 3:2) with a potential 
applied between ring and end-caps of $
V_0 \cos\Omega_Tt + U_0$ with $\Omega_T/2\pi \approx  238$ 
MHz, $V_0 \approx 520$ V.  The
trap is typically operated over the range  
$12$ V $< U_0 < 17$ V leading to secular
frequencies  of 
$(\omega_x,\omega_y,\omega_z)/2\pi = (7.3,16,12.6)$ to $(8.2,17.2,10.1)$
MHz.  The ion-ion spacing (along $\hat x$) is $l \approx 2 \mu$m.

\begin{figure}
\epsfig{file=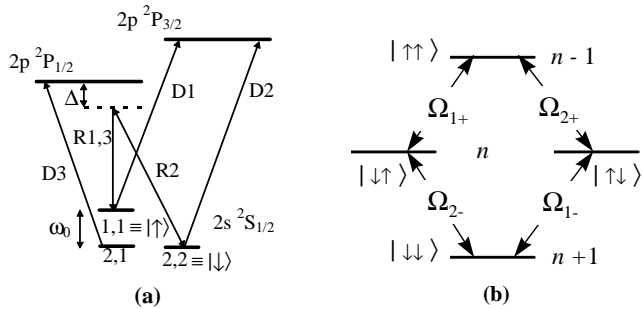,width=3.5in}
\caption{
(a) Relevant $^9$Be$^+$ energy levels.  All optical transitions are near 
$\lambda = 313$ nm, $\Delta/2\pi = 40$ GHz and $\omega_0/2\pi = 1.25$ GHz.
R1-3: Raman beams. 
D1-3: Doppler cooling,
optical pumping and detection beams.
(b) The internal basis qubit states of two spins shown with the vibrational
levels connected on the red motional sideband.  The 
labeled atomic states are as in (a); $n$ is the motional-%
state quantum number (note that the motional mode frequency
$\omega_{\rm str} \ll \omega_0$).  $\Omega_{i\pm}$ are the Rabi frequencies
connecting the states indicated.}
\label{fig:exp}
\end{figure}

 The relevant level structure of $^9$Be$^+$
is shown in Fig.\ \ref{fig:exp}a.  The qubit states
are the $2s\;^2S_{1/2}\;\vert F = 2, m_F = 2 \rangle
\equiv \down$ and
 $2s\;^2S_{1/2}\;\vert F = 1, m_F = 1 \rangle
\equiv \up$ states.  Laser beams  D1 and D2 provide Doppler precooling and 
beam D3 prevents optical pumping to the $\vert F=2,m_F = 1 \rangle$ state.
The cycling $\down \rightarrow 
2p\;^2P_{3/2}\;\vert F = 3, m_F = 3 \rangle$ transition driven by
the $\sigma^+$-polarized D2 laser beam
allows us to differentiate 
$\up$ from $\down$ in a single ion with $\approx$90\%
detection efficiency by observing the fluorescence rate.

Transitions $\down
\ket{n} \leftrightarrow \up \ket{n'}$ (where $n,n'$ are 
vibrational quantum numbers)
are driven by stimulated Raman processes from pairs
of laser beams in one of two geometries.  
Two types of transitions are driven: the ``carrier'' with $n'=n$, and
the red motional sideband (rsb) with $n'=n-1$ \cite{Meekhof96}. 
With reference to Fig. \ref{fig:exp}a, the pair of Raman beams R1 $\perp$ 
R2  has difference-wavevector $\vec{\delta k} 
\parallel \hat x$ and is used for 
sideband cooling (to prepare $\dd\ket{0}$), 
driving the $\hat x$-rsb,
and to drive the ``$\hat x$-carrier''.
Beam pair R2 $\parallel$ R3 
has $\vec{\delta k} \approx 0$ and is not sensitive to motion;
this pair drives the ``co-propagating carrier'' transition.

Two trapped ions aligned along $\hat x$ have two modes of motion
along $\hat x$:  the center-of-mass (COM) 
mode at frequency $\omega_x$ and the stretch mode,
at frequency $\omega_{\rm str} = \sqrt{3}\omega_x$ in which
the two ions move in opposite directions.  We sideband-cool both
of these modes to near the ground state, but use the stretch mode
on transitions which involve the motion since it is
colder (99\% probability of $\ket{n=0}$)
than the COM and heats at a significantly
reduced rate\cite{King98}.  The relevant two-ion qubit level structure
dressed by the quantized harmonic stretch motion is 
shown in Fig.\ \ref{fig:exp}b (we leave out the COM for clarity).  
In general, all four Rabi rates 
$\Omega_{i\pm}$, $i \in \{1,2\}$ connecting the levels
are different and depend on $n$. 
Fig.\ \ref{fig:exp}b shows the states coupled on the rsb
with Rabi frequencies (in the Lamb-Dicke limit)
\begin{equation}
\Omega_{i+} =  \sqrt{n}\;\eta'\Omega_{i} ; \;\;
\Omega_{i-} = \sqrt{n+1}\;\eta'\Omega_{i}  
\label{eq:rsbf}
\end{equation}
where $\eta' = \eta/\sqrt{2\sqrt{3}}$ is the stretch-mode
 two-ion Lamb-Dicke parameter 
(with single-ion  $\eta \approx 0.23$ for 
$\omega_x/2\pi \approx 8$ MHz) and $\Omega_i$ is the carrier
Rabi frequency of ion $i$ \cite{Monroe95}.
On the carrier, the ions are not coupled and the time evolution
is simply that of independent coherent Rabi oscillations
with Rabi frequencies
$\Omega_i$. 
On the co-propagating carrier,
$\Omega_1 = \Omega_2 \equiv \Omega_c$.

In the Cirac-Zoller scheme, each of an array 
of tightly focused laser beams illuminates one and only one ion for
individual state preparation. Here we  
pursue an alternative technique, 
based not on $\Omega_i \rightarrow 0$ for all but one ion,
but simply on $\Omega_1 \neq \Omega_2$.
Differential Rabi frequencies can be used conveniently for individual
addressing on the $\hat x$-carrier:
for example, if 
$\Omega_1 = 2\Omega_2$,
then ion 1 can be driven for a time $\Omega_1t = \pi$ ($2\pi$-pulse,
no spin-flip)
while ion 2 is driven for a $\pi$-pulse resulting in a spin-flip.

\begin{figure}
\epsfig{file=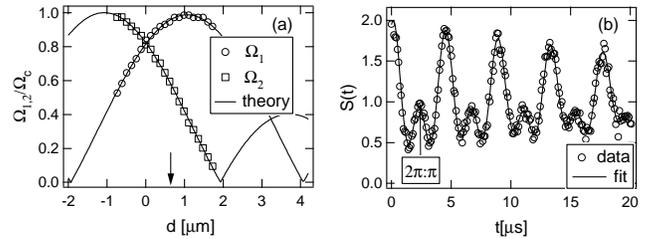,width=3.5in}
\caption{(a) Normalized $\hat x$-carrier Rabi frequencies 
$\Omega_i/\Omega_c$ of each of two ions
as a function of center-of-mass displacement from the rf-null position $d$.  
The solid curves are Eq.\ \protect\ref{eq:micro} where
the distance between the maxima of the two curves 
sets the scale of the ordinate, based on the known ion-ion spacing
of $l \approx 2.2 \mu$m at $\omega_x/2\pi = 8.8$ MHz.
(b) Example of Rabi oscillations with 
$\Omega_1 = 2\Omega_2$. A fit to Eq.\ \protect\ref{eq:S} 
determines that 
$\Omega_1/2\pi = 2\Omega_2/2\pi \approx 225$ kHz,
$\gamma/2\pi \approx 6$ kHz and
$\alpha \approx -0.05$.  The arrow in (a) indicates the conditions of (b).}
\label{fig:Rabis}
\end{figure}

Our technique for differential addressing is to control the ion micro-motion.
To a good approximation, we can write
\begin{equation}
\Omega_i = \Omega_c J_0 (\vert\vec{\delta k}\vert \xi_i)
\label{eq:micro}
\end{equation}
where $J_0$ is the zero-order Bessel function
and $\xi_i$ is the amplitude of micro-motion (along $\hat x$)
associated with ion $i$,
proportional to the ion's mean displacement from trap center. The
micro-motion is controlled by applying a static electric field to push the
ions \cite{Jefferts95}
along $\hat x$, moving ion 2 (ion 1) away from (toward) 
the rf null position, inducing a smaller (larger) Rabi frequency.
The range of Rabi
frequencies explored experimentally is shown in Fig. \ref{fig:Rabis}a.

We determine $\Omega_{1,2}$
by observing the Rabi oscillations of the ions
driven on the $\hat x$-carrier. 
An example with
$\Omega_1 = 2\Omega_2$ is shown in Fig.\ \ref{fig:Rabis}b.
We detect a fluorescence signal
$S(t) = 2P_{\downarrow\downarrow} + (1 + \alpha) P_{\downarrow\uparrow}
+ (1 - \alpha) P_{\uparrow\downarrow}$
where $P_{kl} = \ns{\psi(t)}{kl}$, $k,l \in \{\uparrow,
\downarrow\}$, $\psi (t)$ is the state at time $t$
and $\vert\alpha\vert \ll 1$ describes a small 
differential detection efficiency
due to the induced differential micro-motion.  
Driving on the $\hat x$-carrier for time $t$ starting from $\dd\ket{0}$,
$S(t)$ can be described by
\begin{eqnarray}
S(t) = 1 &+& (1/2)(1+\alpha) \cos (2\Omega_1t) e^{-\gamma t}
\nonumber \\ &+& 
(1/2)(1-\alpha) \cos (2\Omega_2t) e^{-(\Omega_2/\Omega_1)\gamma t}
\label{eq:S}
\end{eqnarray}
where $\gamma$ allows for decay of the signal \cite{Meekhof96}.
The local maximum at $t = 2.4$ $\mu$s on Fig.\ \ref{fig:Rabis}b
is the $2\pi:\pi$ point at which ion 1 has undergone a 
$2\pi$-pulse while ion 2 has undergone a $\pi$-pulse resulting in
$\dd\ket{0} \rightarrow \du\ket{0}$.  Driving a $\pi:\pi$
pulse on the co-propagating carrier transforms 
$\du\ket{0}$ to $\ud\ket{0}$ and $\dd\ket{0}$ to $\uu\ket{0}$, 
completing generation of all four 
internal basis states of Fig.\ \ref{fig:exp}b.

Now consider the 
levels coupled by the first rsb
shown in Fig.\ \ref{fig:exp}b.  If we start in the state 
$\ket{\psi(0)} = \du \ket{0}$ 
and drive on the (stretch mode) rsb for time $t$,
\begin{eqnarray}
\ket{\psi(t)} &=& -\frac{i\Omega_{2-}}{G} \sin(Gt) 
\dd \ket{1} \nonumber \\
&+&\left[ \frac{\Omega_{2-}^2}{G^2}\left( \cos Gt - 1 \right) + 1\right]
\du \ket{0} \nonumber \\ 
&+&e^{i\phi}\left[ \frac{\Omega_{2-} \Omega_{1-}}{G^2} \left(\cos Gt - 1 \right)\right]
\ud \ket{0}
\label{eq:state}
\end{eqnarray}
where
$G = (\Omega_{2-}^2 + \Omega_{1-}^2)^{1/2}$ with $\Omega_{i-}$
from Eq. \ref{eq:rsbf} with $n = 0$.  The phase factor
$\phi = \vec{\delta k} \cdot \langle \vec x_1 - \vec x_2 \rangle$ 
depends on the spatial separation of the ions and arises
because each ion sees a different phase in the $\hat x$ travelling-wave
Raman field. 
The ion-ion spacing
varies by $\delta l\approx 100$ nm over the range of $U_0$ cited above
($\phi = 0$ for $U_0 = 16.3$ V and
$\phi = \pi$ for $U_0 = 12.6$ V, with $d\phi/dU_0$ in good agreement with 
theory). For $Gt = \pi$, the final state is
$\psi_e(\phi)$ from Eq.\ \ref{eq:estate}. 
Note that  $\Omega_1 = (\sqrt{2}+1)\Omega_2$
would generate the Bell states (but we would not
have access to the initial state $\du$, since $\Omega_i$ are
fixed throughout an experiment). 

\begin{figure}
\epsfig{file=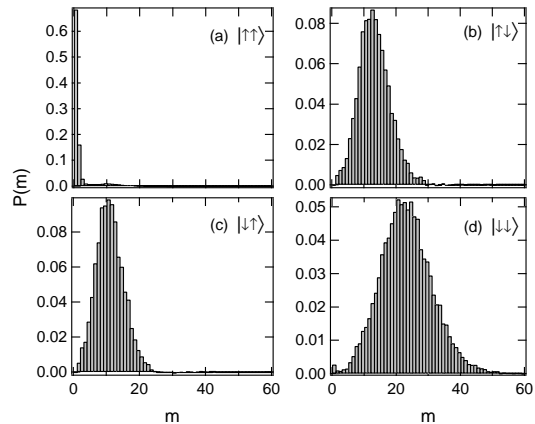,width=3in}
\caption{Photon-number distributions for the four basis qubit states.
Plotted in each graph is the probability of occurrence $P(m)$ of $m$ photons
detected in 500 $\mu$s {\it vs.} $m$, taken over $\sim 10^4$ trials.
Note the different scales for each graph.
}
\label{fig:refs}
\end{figure}

We now describe our two-ion state-detection
procedure.
We first prepare a two-ion basis state $\ket{kl}$, apply the detection
beam D2 for a time $\tau_d \approx 500 \mu$s and
record the number of photons $m$ detected in time $\tau_d$.
We repeat this sequence for $N \approx 10^4$ trials and build a histogram
of the photons collected
(Fig.\ \ref{fig:refs}).
To determine the populations of an unknown state, 
we fit its histogram
to a weighted sum of the four reference histograms
with a simple linear least-squares procedure.

We observe that the $\uu$ 
count distribution (Fig.\ \ref{fig:refs}a) 
is not a single peak at
$m = 0$, corresponding to the expected zero
scattered photons.  Counts at $m = 1$ and $m = 2$ are due to a background 
of 200-400 photons per second. The counts in bins $m > 2$
(which account for $\sim$ 10\% of the area) are due to
a depumping process in which D2
off-resonantly drives an ion out of $\up$
ultimately trapping it in the cycling transition. 
We  approximately double the depumping time 
by applying two additional Raman ``shelving'' pulses
($\up \rightarrow$
$^2S_{1/2}\ket{F=2,m_F=0}\rightarrow$$^2S_{1/2}
 \vert F = 1, m_F = -1\rangle$;
$\down$ unaffected) after every state preparation.
Nevertheless, this results in an average difference of only 10-15
detected photons between an initial $\down$ and $\up$ state,
as shown in Fig.\ \ref{fig:refs} \cite{dp}.   
The distributions associated with $\du$ ,$\ud$ 
and $\dd$ are non-Poissonian 
due to detection laser intensity and frequency fluctuations,
the depumping described above and $\down \rightarrow \up$
transitions from imperfect polarization of D2.

One may ask: what is our overall two-ion state-detection efficiency
on a {\it per experiment} basis?
 To address this issue, we
distinguish three cases:
1) $\uu$, 2) $\ud$ or $\du$, 3) $\dd$.  
Now define case 1 to be true when $m \leq 3$, case 2 when
$3 < m < 17$, and case 3 when $m \geq 17$.
This gives an optimal 80\% probability
that the inferred case (1, 2, or 3)
from a measured $m$ in a single experiment is the actual case.

\begin{figure}
\epsfig{file=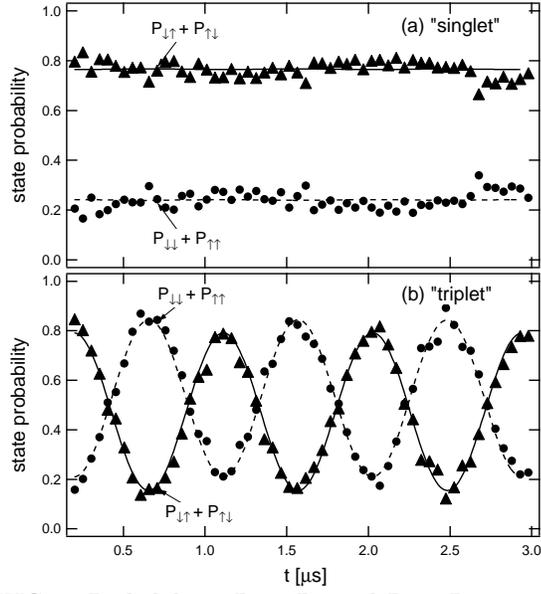,width=3in}
\caption{Probabilities 
$P_{\downarrow\uparrow} + P_{\uparrow\downarrow}$
and
$P_{\downarrow\downarrow} + P_{\uparrow\uparrow}$
as a function of time $t$ driving on the
co-propagating carrier, starting from
(a) the ``singlet'' $\psi_e(0)$ and
(b) the ``triplet'' $\psi_e(\pi)$ entangled states.
The equivalent rotation angle is $2\Omega_c t$
($\Omega_c/2\pi \approx 200$ kHz for these data).
The solid and dashed lines in (a) and (b) are sinusoidal fits
to the data, from which the contrast is extracted.
}
\label{fig:evolve}
\end{figure}

We have generated states described by
density operators $\rho^\pm$ in which the populations (diagonals
of $\rho^\pm$) are measured to be
$P_{\downarrow\uparrow} \approx P_{\uparrow\downarrow} 
\approx 0.4, \; P_{\downarrow\downarrow} \approx 0.15, \;
P_{\uparrow\uparrow} \approx 0.05$.
To establish coherence, consider first the 
Bell singlet state $\psi_B^-$
which has $P_{\downarrow\uparrow} = P_{\uparrow\downarrow} = 1/2$.
Since $\psi_B^-$ has
total spin $J = 0$, any $J$-preserving transformation,
such as an an equal rotation on both spins, must leave
this state unchanged, whereas such a rotation on a mixed state with
populations $P_{\downarrow\uparrow} = P_{\uparrow\downarrow} = 1/2$
and no coherences will evolve quite differently.
We rotate both spins trough an angle $\theta$ 
by driving on the co-propagating carrier for a time $t$ such that
$\theta = \Omega_c t$.  
Fig. \ref{fig:evolve}a shows the time evolution of 
an experimental state which approximates the singlet Bell state. 
Contrast this with the approximate ``triplet'' state 
shown in Fig. \ref{fig:evolve}b.  
More quantitatively, the data show that $\rho^\pm$ is
decomposed as
$\rho^\pm = C\vert\psi_B^\pm\rangle\langle\psi_B^\pm\vert + (1-C)\rho_m$
in which $\rho_m$ has no coherences 
which contribute to the measured signal
(off diagonal elements
connecting $\ud$ with $\du$ and $\uu$ with $\dd$), and
$C=0.6$ is the contrast of the curves in
Fig.\ \ref{fig:evolve}. This leads to a fidelity of
$\langle\psi_B^\pm\vert\rho^\pm\vert\psi_B^\pm\rangle =
(P_{\downarrow\uparrow} + P_{\uparrow\downarrow} + C)/2 \approx 0.7$.

The non-unit fidelity of our states arises from 
several technical factors. The first is Raman
laser intensity noise which gives rise to a
noise-modulated Rabi-frequency.  The second is a second-order (in $\eta$)
effect on $\Omega_i$ due to the motional state of the
COM mode \cite{Wineland98}, 
which is not in the ground state at all times \cite{King98}.
These effects can be seen in Fig.\ \ref{fig:Rabis}b as a decay
envelope on the data (modeled by $\gamma$ of Eq.\ \ref{eq:S})
and cause a 10\% infidelity in initial state preparation \cite{infidel}.
This initial imperfection in
state preparation, the contribution of the above factors on the rsb
pulse and a first order effect due to imperfect ground-state
preparation of the stretch mode
are responsible for the rest of the infidelity.

The micro-motion-induced selection of Rabi frequencies
as here demonstrated is sufficient to implement
universal quantum logic with individual addressing \cite{Cirac95}. 
 To isolate ion 1,
we arrange the trap strength
and static electric field so that ion 1 is at the rf null position 
($\Omega_1 = \Omega_c J_0(0) = \Omega_c$) and ion 2
is at a position such that 
$\Omega_2 = \Omega_c J_0(\vert\vec{\delta k}\vert\xi_2) = 0$. To isolate 
ion 2, we drive on the first motional sideband of the
rf-micro-motion by adding
$\Omega_T/2\pi = \pm 238$ MHz to the difference frequency 
of the Raman beams resulting in 
$\Omega_1 = \Omega_c J_1(0) = 0$ and $\Omega_2 = 
\Omega_c J_1(\vert\vec{\delta k}\vert\xi_2)=
0.519 \Omega_c$.  
This provides a means of individual addressing, 
with which the Cirac-Zoller scheme \cite{Cirac95} can be implemented
for two ions.

In conclusion, we have taken a first step in the quantum preparation
and manipulation of entangled states of multiple trapped ions--- a step 
which is crucial for quantum computations with trapped ions.  We have
{\it engineered} entangled states deterministically, that is, there is no
inherent 
probabilistic nature to our quantum entangling source.  We have developed a
two-ion state-sensitive detection technique which allows us to measure 
the diagonal elements of the density matrix $\rho^\pm$ of our states,
 and have performed
transformations which directly measure the relevant off-diagonal coherences
of $\rho^\pm$.

We acknowledge support from the U.\ S.\ National Security Agency, Office of 
Naval Research and Army Research Office. We thank Eric Cornell, Tom Heavnor,
David Kielpinski, and Matt Young for critical readings of the manuscript.

%\enlargethispage{10\baselineskip}

%\enlargethispage{10\baselineskip}

\begin{references}
\bibitem[*]{gov} Work of the US government.  Not subject to US copyright.
\bibitem{dp}
This off-resonant de-pumping can be suppressed
by increasing the energy separation between $\down$ and $\up$
with a strong magnetic field or by using a heavier ion with larger
hyperfine splitting.
\bibitem{infidel}
The reference histograms for $\du$ and $\ud$ 
(Figs.\ \ref{fig:refs}b,c) have had this 10\% contamination
from $\uu$ and $\dd$ removed, in order to assure
that the references to which we fit are accurate representations
of the underlying states.
\end{references}
\end{document}